\documentclass[ 
 reprint,
 amsmath,amssymb,
 aps,
floatfix,
]{revtex4-2}

\usepackage{xcolor}
\usepackage{placeins}
\usepackage{graphicx}
\usepackage{dcolumn}
\usepackage{bm}

\usepackage{graphicx}
\usepackage{dcolumn}
\usepackage{bm}

\usepackage[utf8]{inputenc}
\usepackage[T1]{fontenc}
\usepackage{mathptmx}
\usepackage{tikz,caption}
\usepackage{enumerate}

\newcommand{\bx}{\mathbf{x}}
\newcommand{\bk}{\mathbf{k}}

\begin{document}


\title{A Scanning 2-Grating Free Electron Mach-Zehnder Interferometer}

\author{Cameron W. Johnson}%
\thanks{These two authors contributed equally}
\email{cwj@uoregon.edu}

\author{Amy E. Turner}
\thanks{These two authors contributed equally}
\author{Benjamin J. McMorran}%
\affiliation{%
 Department of Physics, University of Oregon, Eugene, Oregon 97403, USA
}

\date{\today}

\begin{abstract}
We demonstrate a 2-grating free electron Mach-Zehnder interferometer constructed in a transmission electron microscope. A symmetric binary phase grating and condenser lens system forms two spatially separated, focused probes at the sample which can be scanned while maintaining alignment. The two paths interfere at a second grating, creating in constructive or destructive interference in output beams. This interferometer has many notable features: positionable probe beams, large path separations relative to beam width, continuously tunable relative phase between paths, and real-time phase information. Here we use the electron interferometer to measure the relative phase shifts imparted to the electron probes by electrostatic potentials as well as a demonstration of quantitative nanoscale phase imaging of a polystyrene latex nanoparticle.
\end{abstract}

{
\let\clearpage\relax
\maketitle
}


\section{Introduction} 

Electron holography and interferometry can enable nanoscale phase imaging \cite{tonomura_electron_1999,dunin-borkowski_electron_2019}, the exploration of the Aharonov-Bohm effect \cite{tonomura_evidence_1986,batelaan_aharonovbohm_2009}, interaction-free measurements and quantum electron microscopy \cite{amy_2021,putnam_noninvasive_2009,kruit_designs_2016}, the measurement of coherence properties \cite{potapov_inelastic_2007,pooch_coherent_2018,kerker_quantum_2020}, quantum state tomography  \cite{priebe_attosecond_2017,madan_holographic_2019}, and the coherent control of the free electron wavefunction \cite{echternkamp_ramsey-type_2016}. While interferometry is widely used in optics and photonics fields such as astronomy \cite{monnier_optical_2003}, optical metrology \cite{groot_review_2019}, and quantum optics \cite{pan_multiphoton_2012}, electron interferometry has advanced at a slower pace, partially due to a lack of basic optical elements such as beamsplitters and mirrors that can be used to build a versatile system. Here we use two diffraction gratings as beamsplitters in a conventional transmission electron microscope (TEM) to create a flexible, path-separated Mach-Zehnder interferometer for free electrons, emulating the canonical example for phase sensitive interferometry.

Electron interferometers are currently limited and defined by the electron optical elements used to construct the interferometer. Electrostatic mirrors for free electrons show great promise, but are in their early development \cite{krielaart_grating_2018,abedzadeh_electrostatic_2020,krielaart_flat_2021}. Beamsplitters have existed for decades, the most prevalent being electrostatic biprisms \cite{lichte_electron_2007} which divide wavefronts and require high spatial coherence \cite{chang_optimising_2015}. Biprisms generate interferograms from which the phase of the sample can be extracted after image processing. However, they require high spatial coherence; thus, they are inherently limited by modern electron emission sources that are only partially coherent. Using recently improved diffraction gratings as amplitude dividing beamsplitters \cite{harvey_efficient_2014,freimund_observation_2001,feist_high-purity_2020} to create spatially separated paths evades the high spatial coherence requirement, if the separated paths are perfectly recombined after the sample \cite{yasin_path-separated_2018}. We note that amplitude division with microwave chip  beamsplitters has also been demonstrated, but is not yet practical for electrons with kinetic energies above 200 eV \cite{zimmermann_beam_2019}.

Various demonstrations of matterwave interferometry have advanced for decades \cite{marton_electron_1952}. Path-separated Mach-Zehnder interferometers specifically have been demonstrated for different kinds of matterwaves, including neutrons \cite{rauch_neutron_2015}, atoms \cite{cronin_optics_2009}, BECs \cite{berrada_integrated_2013}, conduction and quantum Hall valley electrons in 2D devices \cite{ji_electronic_2003,jo_quantum_2021}, and SQUIDS \cite{oliver_mach-zehnder_2005}. Free electron Mach-Zehnder interferometers with discrete outputs have predominantly been constructed using nanofabricated diffraction gratings or crystals as beamsplitters. Two-plane electron interferometers fabricated from monolithic uniform crystals \cite{tavabi_new_2017,agarwal_nanofabricated_2017} and three-plane Mach-Zehnder interferometers with discrete crystal planes \cite{marton_electron_1952,marton_electron_1954} have been demonstrated. A three-grating Mach-Zehnder interferometer for free electrons was demonstrated with nanofabricated transmission amplitude gratings \cite{gronniger_three-grating_2006}. Even though all of these interferometer variants for free electrons have advanced electron interferometry, none of them have the ability to scan the probes over a specimen; i.e., they cannot be used for imaging. Furthermore, the short lengthscales of these interferometers has restricted the type of experiments that can be performed.

Here we demonstrate a 2-grating electron Mach-Zehnder interferometer (2GeMZI) insided a conventional TEM that provides clearly defined isolated probes, arbitrary probe phase shifts, and scanning/imaging capabilities. Furthermore, since this is a scanning probe technique, the magnification can be changed at will without adjusting the setup, unlike electron holography. This is accomplished by placing diffraction gratings in apertures above and below the specimen plane of a TEM operated in scanning TEM (STEM) mode. The small deviations of the lens and aperture positions from the nominal STEM settings allow us to maintain the high resolution imaging capabilities afforded by the TEM while retaining the precise interferometer alignments. To demonstrate the phase sensitivity of the 2GeMZI we map electrostatic potential differences in the vicinity of both grounded and charged silver nanorods and demonstrate quantitative nanoscale phase imaging of a spherical latex nanoparticle. 

\section{Theoretical Description of Electron Interferometer} 

In this electron interferometer, nanoscale diffraction gratings are used as amplitude-dividing beamsplitters and the standard TEM imaging optics are used to separate, scan, and recombine the beams. An illustration of this system can be seen in Figure \ref{fig:fig1}(a) where successive transverse planes are defined in relation to the previous plane by Fourier transform. Here we describe how the evolution of the electron wavefunction can be modeled throughout the interferometer.

The evolution of an electron wavefunction propagating through free space, neglecting spin, can be described by the Schr\"odinger equation with relativistic corrections. In a typical TEM with a field emission electron source, electrons are accelerated to beam energies of 40 to 300 keV with a 500 meV energy spread and the electron beam has very small beam divergence \cite{reimer_transmission_2013}. The evolution of electron wavefunctions can therefore be modeled using Fresnel and Fourier optical theory \cite{goodman_introduction_2005}, consistent with the Schr\"odinger equation with assumptions that the electrons are largely quasimonochromatic, non-interacting, and collimated.

An electron wavefunction passing near or transmitting through an object can accumulate phase shifts and amplitude losses. In the weak phase approximation \cite{cowley_electron_1972}, these effects are proportional to the longitudinal extent of the interaction, e.g., the thickness of the material. This can be described by a complex index of refraction $\eta=\sigma V_{\text{mip}}+i\gamma$, where $\sigma = 2\pi me\lambda/h^2$ is the object independent interaction parameter for a free electron with relativistic mass $m$ and de Broglie wavelength $\lambda$, $V_{\text{mip}}$ is the mean inner potential of the material, and $\gamma$ is a material-dependent decay coefficient that models coherent amplitude loss due to high-angle scattering. We use this complex index of refraction model to describe both the diffraction holograms we employ as beamsplitters and the specimens we image.

The transverse electron wavefunction incident on the input grating of the interferometer is assumed to be a plane wave with an outer edge defined by an aperture $\psi^{\{1\}}_{in}(\bx)=A(\bx)$. When transmitted through a grating the wavefunction is modified by its transmission function 
\begin{equation}
    \psi^{(1)}_{out}(\bx) = A(\bx)e^{i\eta t_1(\bx)}, \label{eq:psi1out}
\end{equation}
where $t_1(\bx)$ is the periodic thickness profile of the grating. For a straight grating with pitch $p_1$ and diffraction wavevector $\bk_1=2\pi/p_1$, we can expand Eq. (\ref{eq:psi1out}) by the Fourier series representation of the exponential term
\begin{equation}
    \psi^{(1)}_{out}(\bx) = A(\bx)\sum_nc^{\{1\}}_ne^{in\bk_1\cdot\bx}, \label{eq:psi1out0}
\end{equation}
where the Fourier coefficients are given by 
\begin{equation}
    c^{\{1\}}_n = \frac{1}{p_1}\int_0^{p_1}d\tilde{x}\,e^{i\eta t_1(\tilde{x})-in|\bk_1|\tilde{x}}
\end{equation}
where $\tilde{x}$ is the 1d direction of the grating pitch.
Then the unnormalized probes in the back focal plane of the input grating can be expressed as 
\begin{equation}
    \begin{split}
        \psi^{(2)}(\bk) \propto & \,\,  \sum_nc^{\{1\}}_n\widetilde{A}(\bk-n\bk_1),\label{eq:psi2}
    \end{split}
\end{equation}
where $\widetilde{A}$ is the Fourier transformed $A$. If the second grating is allowed to translate by an amount $\bx_0'$, then the output of the second grating is
\begin{equation}
    \psi^{(3)}_{out}(\bx') \propto A(\bx')\sum_{n,n'}c_n^{\{1\}*}c^{\{2\}}_{n'}e^{in'\bk_2\cdot(\bx'-\bx_0')-in\bk_1\cdot\bx'}, \label{eq:psi2out}
\end{equation}
where $c_n^{\{2\}}$ and $\bk_2$ are similarly defined for the second grating. When the image of the input grating is projected onto the output grating with the same pitch and orientation, i.e., $\bk_0 = \bk_1 = \bk_2$, the wavefunction in final $\bk'$ plane can be written as
\begin{equation}
    \psi^{(4)}(\bk') \propto \sum_{n,n'}c_n^{\{1\}*}c_{n'}^{\{2\}}e^{-in'\bk_0\cdot\bx_0'}\widetilde{A}(\bk'-(n-n')\bk_0).\label{eq:psi4}
\end{equation}
The output of the interferometer is divided into distinct $m=n-n'$ diffraction orders. Using this to re-index the double sum, we can write Eq. (\ref{eq:psi4}) as a sum of output diffraction orders 
\begin{equation}
    \begin{split}
        \psi^{(4)}(\bk') \propto & \,\, \sum_{m}\left(\sum_nc_n^{\{1\}*}c_{n-m}^{\{2\}}e^{-i(n-m)\bk_0\cdot\bx_0'}\widetilde{A}(\bk'-m\bk_0)\right) \\
        = & \,\, \sum_{m}\psi_m^{(4)}(\bk').\label{eq:psi40}
    \end{split}
\end{equation}When the gratings are symmetric and put a majority of the the transmitted intensity into the $\pm1^{\text{st}}$ diffraction orders, i.e., $|c_{\pm1}|\gg|c_{|n|\neq1}|$ and $|c_{+1}|=|c_{-1}|$, then the $0^{\text{th}}$ output diffraction order where $n=n'$, up to a global phase, is 
\begin{equation}
    \begin{split}
        \psi^{(4)}_{0}(\bk') \propto & \,\, |c_1^{\{1\}}c_1^{\{2\}}|\left(1+e^{-2i\bk_0\cdot\bx_0'}\right)\widetilde{A}(\bk') + \cdots.  \label{eq:psi41}
    \end{split}
\end{equation}

If the probes are allowed to scan in the specimen plane, $\bk\rightarrow\bk+\bk_s$, while passing through some electrostatic potential $V(\bk,z)$, we can use the weak phase approximation and write the phase accumulated by each probe as being proportional to the projected potential along the $z$ direction $\Phi(\bk) = \sigma \int dz\,V(\bk,z)$.
With this the $0^{\text{th}}$ order output of the interferometer is approximately
\begin{equation}
    \begin{split}
        \psi^{(4)}_{0}(\bk') \propto & \,\, \left(1+e^{i\varphi(\bk_s,\bx'_0)}\right)\widetilde{A}(\bk'), \label{eq:phi42} 
    \end{split}
\end{equation}
where the total phase difference between the two highest intensity probes is 
\begin{equation}
    \begin{split}
        \varphi(\bk_s,\bx'_0)=& \,\,-2\bk_0\cdot\bx_0' +\Phi(\bk_s-\bk_0)-\Phi(\bk_s+\bk_0). \label{eq:dphs}
    \end{split}
\end{equation}
We should note that the static potential $V$ can have contributions that extend into the vacuum region due to the build up of surface charge on a material as well as inside materials from the mean inner potential as was defined in $\eta$. 


\onecolumngrid
\begin{center}
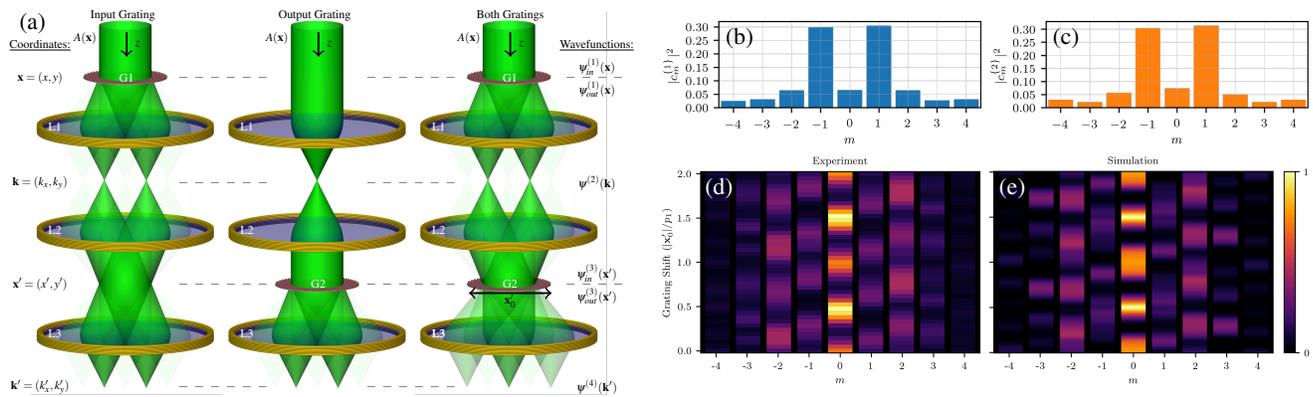
\begin{figure}[h]
\begin{tikzpicture}
    \node[inner sep=0pt] at (0,0)
        {\includegraphics[width=7in]{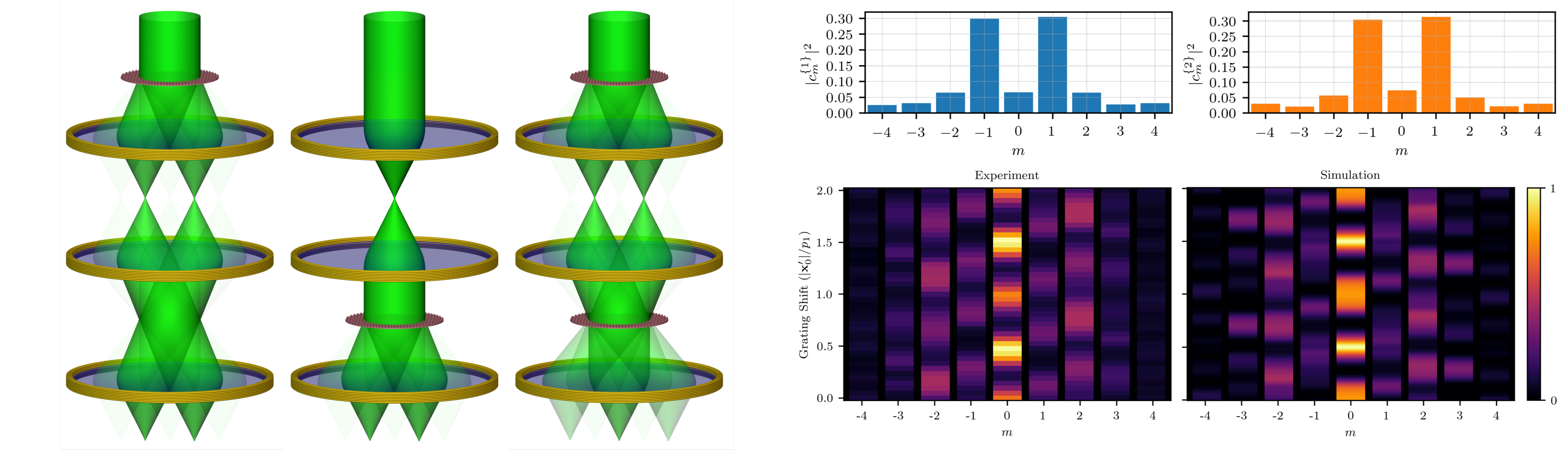}};
    \draw[dashed,color=gray,line width=0.125mm](-6.25,1.675)--(-5,1.675);
    \draw[dashed,color=gray,line width=0.125mm](-6.25,0.275)--(-5,0.275);
    \draw[dashed,color=gray,line width=0.125mm](-6.25,-1.075)--(-5,-1.075);
    \draw[dashed,color=gray,line width=0.125mm](-6.25,-2.44)--(-5,-2.44);
    \draw[dashed,color=gray,line width=0.125mm](-3.75,1.675)--(-2.5,1.675);
    \draw[dashed,color=gray,line width=0.125mm](-3.75,0.275)--(-2.5,0.275);
    \draw[dashed,color=gray,line width=0.125mm](-3.75,-1.075)--(-2.5,-1.075);
    \draw[dashed,color=gray,line width=0.125mm](-3.75,-2.44)--(-2.5,-2.44);
    \node[scale=0.5,inner sep=0pt] at (-2.8,1.025)
        {\textcolor{white}{L1}};
    \node[scale=0.5,inner sep=0pt] at (-2.8,-0.35)
        {\textcolor{white}{L2}};
    \node[scale=0.5,inner sep=0pt] at (-2.8,-1.725)
        {\textcolor{white}{L3}};
    \node[scale=0.5,inner sep=0pt] at (-5.35,1.025)
        {\textcolor{white}{L1}};
    \node[scale=0.5,inner sep=0pt] at (-5.35,-0.35)
        {\textcolor{white}{L2}};
    \node[scale=0.5,inner sep=0pt] at (-5.35,-1.725)
        {\textcolor{white}{L3}};
    \node[scale=0.5,inner sep=0pt] at (-2.8,-1.725)
        {\textcolor{white}{L3}};
    \node[scale=0.5,inner sep=0pt] at (-7.9,1.025)
        {\textcolor{white}{L1}};
    \node[scale=0.5,inner sep=0pt] at (-7.9,-0.35)
        {\textcolor{white}{L2}};
    \node[scale=0.5,inner sep=0pt] at (-7.9,-1.725)
        {\textcolor{white}{L3}};
    \node[scale=0.5,inner sep=0pt] at (-6.95,1.675)
        {\textcolor{white}{G1}};
    \node[scale=0.5,inner sep=0pt] at (-1.85,1.675)
        {\textcolor{white}{G1}};
    \node[scale=0.5,inner sep=0pt] at (-1.85,-1.07)
        {\textcolor{white}{G2}};
    \node[scale=0.5,inner sep=0pt] at (-4.4,-1.07)
        {\textcolor{white}{G2}};
    \node[scale=0.5,inner sep=0pt] at (-7.5,2.2)
        {\textcolor{black}{$A(\bx)$}};
    \node[scale=0.5,inner sep=0pt] at (-4.95,2.2)
        {\textcolor{black}{$A(\bx)$}};
    \node[scale=0.5,inner sep=0pt] at (-2.4,2.2)
        {\textcolor{black}{$A(\bx)$}};
    \node[scale=0.5,inner sep=0pt] at (-1.7,1.975+0.15) {$z$};
    \draw[->, line width=0.25mm] (-1.85,2.125+0.15) -- (-1.85,1.8+0.15);
    \node[scale=0.5,inner sep=0pt] at (-4.25,1.975+0.15) {$z$};
    \draw[->, line width=0.25mm] (-4.4,2.125+0.15) -- (-4.4,1.8+0.15);
    \node[scale=0.5,inner sep=0pt] at (-6.8,1.975+0.15) {$z$};
    \draw[->, line width=0.25mm] (-6.95,2.125+0.15) -- (-6.95,1.8+0.15);
    \node[scale=0.5,inner sep=0pt] at (-8.1,2.1)
        {\underline{Coordinates:}};
    \node[scale=0.5,inner sep=0pt] at (-8.1,1.675)
        {$\bx=(x,y)$};
    \node[scale=0.5,inner sep=0pt] at (-8.1,0.275)
        {$\bk=(k_x,k_y)$};
    \node[scale=0.5,inner sep=0pt] at (-8.1,-1.075)
        {$\bx'=(x',y')$};
    \node[scale=0.5,inner sep=0pt] at (-8.1,-2.44)
        {$\bk'=(k_x',k_y')$};
    \node[scale=0.5,inner sep=0pt] at (-0.7,2.1)
        {\underline{Wavefunctions:}};
    \node[scale=0.5,inner sep=0pt] at (-0.7,1.825)
        {$\psi^{(1)}_{in}(\bx)$};
    \draw[dashed,color=gray,line width=0.125mm](-1.1,1.675)--(-0.3,1.675);
    \node[scale=0.5,inner sep=0pt] at (-0.7,1.525)
        {$\psi^{(1)}_{out}(\bx)$};
    \node[scale=0.5,inner sep=0pt] at (-0.7,0.275)
        {$\psi^{(2)}(\bk)$};
    \node[scale=0.5,inner sep=0pt] at (-0.7,-0.925)
        {$\psi^{(3)}_{in}(\bx')$};
    \draw[dashed,color=gray,line width=0.125mm](-1.1,-1.075)--(-0.3,-1.075);
    \node[scale=0.5,inner sep=0pt] at (-0.7,-1.225)
        {$\psi^{(3)}_{out}(\bx')$};
    \node[scale=0.5,inner sep=0pt] at (-0.7,-2.44)
        {$\psi^{(4)}(\bk')$};
    \node[inner sep=0pt] at (-8.2,2.4)
        {(a)};
    \node[inner sep=0pt] at (1.2,2.2)
        {(b)};
    \node[inner sep=0pt] at (5.54,2.2)
        {(c)};
    \node[inner sep=0pt] at (0.925,0.2)
        {\textcolor{white}{(d)}};
    \node[inner sep=0pt] at (4.825,0.2)
        {\textcolor{white}{(e)}};
    \node[scale=0.5,inner sep=0pt] at (-1.85,-1.3) {$\bx_0'$};
    \draw[<->, line width=0.25mm] (-1.3,-1.195) -- (-2.4,-1.195);
    \node[scale=0.5,inner sep=0pt] at (-7,2.5)
        {Input Grating};
    \node[scale=0.5,inner sep=0pt] at (-4.45,2.5)
        {Output Grating};
    \node[scale=0.5,inner sep=0pt] at (-1.85,2.5)
        {Both Gratings};
\end{tikzpicture}
\caption{\label{fig:fig1}(a) Diagram for 2GeMZI showing definitions of the different transverse planes as well as labels for the transverse wavefunctions in each plane, the magnetic lenses (L1, L2, L3), gratings (G1, G2), and beam-defining aperture $A(\bx)$ are also shown. The three different images depict the 3 cases of when the different gratings are inserted or removed. The position of the second grating relative to the beam is denoted as $\bx_0'$. (b) Diffraction efficiencies of G1, measured when G2 is removed. (c) Diffraction efficiencies of G2, measured when G1 is removed. (d) Measured output intensities of the 2GeMZI for different relative grating shifts normalized to maximum output intensity. (e) Simulated output intensities of the 2GeMZI for different relative grating shifts. }
\end{figure}
\end{center}
\FloatBarrier
\twocolumngrid

\section{Construction of Interferometer}

Two arrays of 350 nm pitch, 30 $\mu$m diameter binary phase gratings were each nanofabricated onto a 250x250 $\mu$m$^2$, 30 nm thick, free-standing Si$_3$N$_4$ membrane using focused ion beam (FIB) gas-assisted etching \cite{johnson_improved_2020}. To aid with alignment, the gratings were patterned at multiple orientations in a 6x6 array. In an image-corrected 80-300 keV FEI Titan TEM, one grating array was installed in the second condenser aperture and used as the initial beamsplitter (G1). The second grating array was installed in the post-specimen selected area aperture as the beam-combining beamspliter (G2). The TEM was operated at 80 keV in STEM mode with approximately a 1 mrad convergence angle. An independently positionable circular aperture at the third condenser lens was used to select the output of a single grating out of the widely illuminated array of gratings. The diffraction orders created by the selected input grating were focused to narrow probes at the sample, the $\bk$ plane. The lenses were set in free lens control with assistance of the "Lorentz" lens in the image corrector to form a correctly magnified, oriented, and in-focus image of G1 onto G2. Finally, the post-G2 projection lenses were used to image the far-field diffraction pattern from both beamsplitters onto the detectors at the bottom of the TEM column. The relative grating shift parameter $\bx_0'$ was controlled by the diffraction alignment coils in the image corrector that shifted the image of G1 relative to G2 allowing for arbitrary relative phase shifts between the two specimen plane probes in the interferometer output. The $\bk$ plane probes at the specimen section could be set to have up to 1 $\mu$m separation between the $\pm1$ probes, although the spot size increases proportionally with the probe separation. Using these beamsplitter gratings, the ratio between separation between the paths and the width of the probes is always about 20. 

The magnitude of the $\{c_n^{\{1\}}\}$ and $\{c_n^{\{2\}}\}$ Fourier coefficients were measured by inserting only one of G1 or G2 at time while collecting an image of the probe intensities with a scintillator fiber-coupled CCD. A single diffraction order was integrated and divided by the total integrated intensity to determine normalized diffraction efficiencies $|c_n^{\{1\}}|^2$ and $|c_n^{\{2\}}|^2$; the measured grating outputs are shown in Figure \ref{fig:fig1}(b,c). Ideally, the gratings would be perfect binary gratings with 50\% groove duty cycle with up to 40.5\% of the transmitted intensity going into the $\pm1$ probes. However, edge rounding and non-ideal duty cycles from nanofabrication with a finite width ion beam as well as over and under milling from the ideal groove depth caused deviations from the optimal diffraction efficiency. Even so, we were able to achieve dominant $\pm1$ coefficients allowing for efficient 2 beam scanning in the 2GeMZI. 
    
With G1 and G2 both inserted, we collected CCD images of the output beams for different values of $\bx_0'$ by tuning the previously mentioned diffraction alignment coils. The measured output intensities are in good agreement with the expected result, Fig. 1(d,e). Without the presence of an electrostatic potential, the intensity of the 0$^{\text{th}}$ order interferometer output is expected to be proportional to the modulus square of Equation (\ref{eq:phi42}), i.e., sinusoidal in the argument $\bk_0\cdot\bx_0'$. We see this dependence in the experimental data, but it is also accompanied by a beating at half the spatial frequency  $\bk_0/2$. This frequency beating is caused by a combination of grating duty cycle mismatch and contributions of the higher order terms from the sum of probe coefficients. Due to these higher order effects, the fringe visibility $\mathcal{V} = (I_{max}-I_{min})/(I_{max}+I_{min})$ is $\mathcal{V}=0.76$ when the output is aligned for maximally destructive interference or $\mathcal{V}=0.82$ when aligned for maximally constructive interference. The maximum theoretical fringe visibility $\mathcal{V}=1$ can be approached through the continued improvement of gratings.

Scan and descan coils can be used to raster both beams across a scan region up to 3 times the probe separation while keeping the electron interference pattern (the image of G1) stationary on the second beamsplitter (G2), ensuring the interferometer output was constant while beams were scanned across a flat phase region. The scan/descan system is independent of the diffraction alignment used to control $\bx_0'$, so the relative phases between the interferometer probes remains constant throughout the scan. It should be noted that phase shifts due to path length differences during the scan certainly exist, but are small enough to neglect. While scanning, a bright field (BF) monolithic detector can be inserted such that it is illuminated by only the 0$^\text{th}$ interferometer output order. With this configuration we can perform direct phase imaging with the 2GeMZI. The relative phase between the probes at any scan position in the specimen plane $\bk_s$ can be reconstructed from the intensity of the interferometer output 
\begin{equation}
    I^{(4)}_0(\bk_s,\bx_0') \approx \langle I^{(4)}_0\rangle \left[1+\mathcal{V}\cos\left(\varphi(\bk_s,\bx_0')\right)\right]. \label{eq:intout}
\end{equation}


\section{Electrostatic Potentials in the Interferometer}

One application of the 2GeMZI is mapping electrostatic potentials in real-time. In the last 30 years, quantitative potential maps measured with electron holography have been used to accurately determine charge distributions of nanoscale devices \cite{matteucci_electron_1992,migunov_model-independent_2015}. However streamlined this method has become, it still requires image post-processing or proprietary live analysis software \cite{mccartney_electron_2007,voelkl_live_2019}. While the high spatial and phase resolution of electron holography cannot yet be matched by the 2GeMZI in this initial demonstration, the interferometer provides a live interpretation of the electrostatic potential, whereas electron holography requires post-scan image processing. Each pixel’s intensity indicates the electrostatic potential difference at the two probe positions in the specimen plane. Here we use the 2GeMZI to show the fringes in raw interferometric BF images of a grounded and insulated vertical nanorod.  

\begin{center}
\begin{figure}[h]
\centering
\begin{tikzpicture}
    \node[inner sep=0pt] at (0,0) {\includegraphics[width=2.7in]{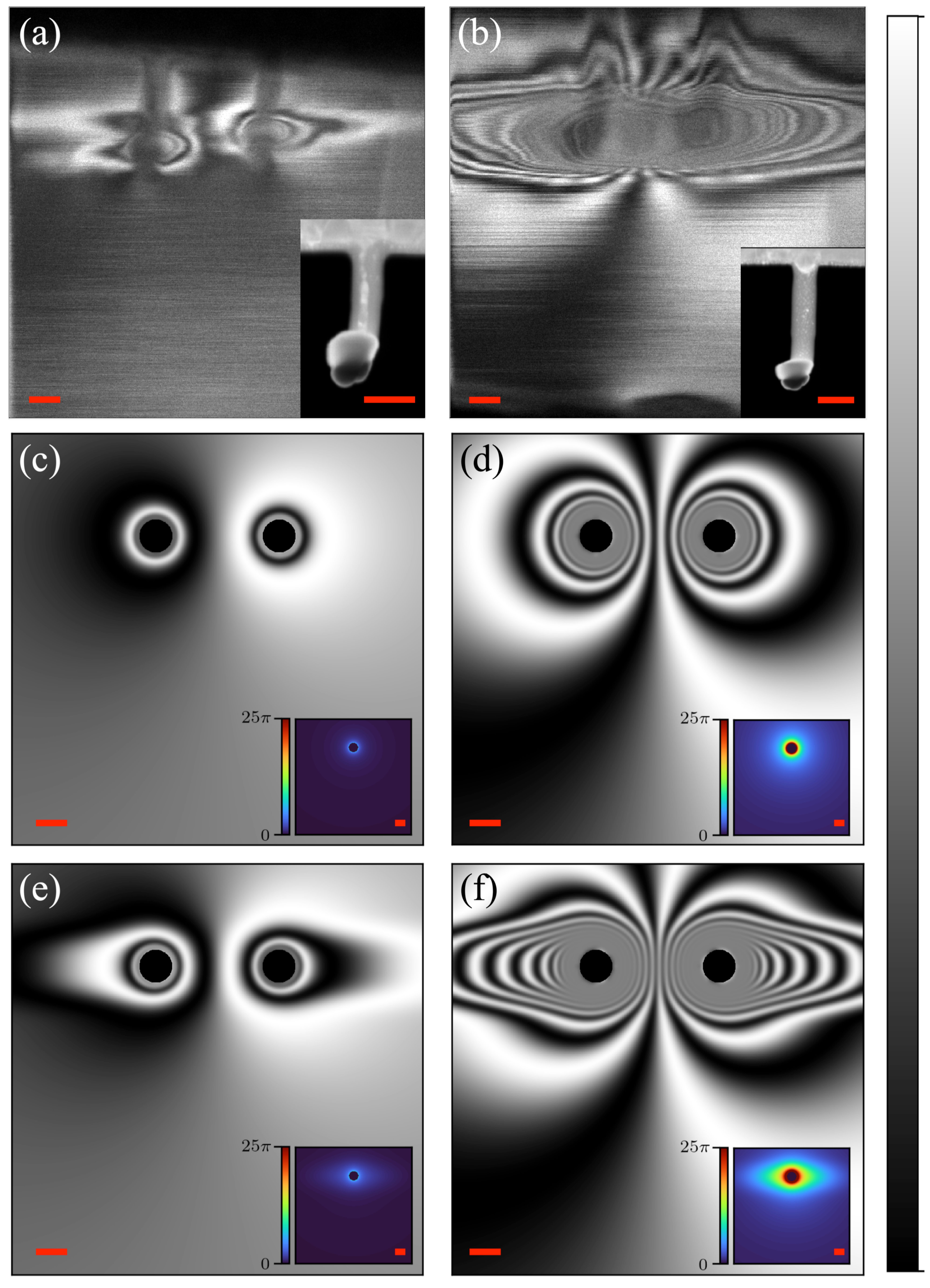}};
    \node[inner sep=0pt] at (3.7,-4.65) {$I_{min}$};
    \node[inner sep=0pt] at (3.7,4.58) {$I_{max}$};
\end{tikzpicture}
\caption{\label{fig:fig2} Bright field (BF) STEM images showing the interferometer output spatially oscillating between the intensity minimum and maximum for (a) grounded vertical Ag nanorod, (b) insulated vertical Ag nanorod. Insets are STEM HAADF images of each nanorod. (c,d) Simulated 2GeMZI output for 2 probes passing through a $1/r$ electrostatic potential where (d) has 10 times the charge of (c). (e,f) Same as (c,d) but with a horizontally elongated Gaussian included with the potential to simulate the increased induced charging. All scale bars are 100 nm.}
\end{figure}
\end{center}

We fabricated two 550 nm tall vertical silver nanorods surrounded by a vacuum window; one with a conductive lead to ground and the other electrically insulated such that it would support a surface charge throughout an imaging scan (see Appendix A for the sample fabrication details). This device was inserted in the specimen plane of the 2GeMZI which was adjusted to have a larger path separation of 500 nm with a probe size of about 25 nm. We recorded interferometric BF images over a scan region of 1.5x1.5 $\mu$m$^2$ of the grounded and insulated nanorods shown in Fig. \ref{fig:fig2}(a,b). The electric potential from the semiconductor nitride substrate cantilever and the grounded nanorod was small; as shown in Fig. \ref{fig:fig2}(a), the interferometer output was only modulated close to the surface of the nitride and the relative phase between two probes in the vacuum region far away from the object is constant. However, the electrically insulated nanorod charged when exposed to the beam until reaching a static surface charge, creating a larger static potential. The resulting interference fringes for the probe potential differences can clearly be seen far into the surrounding vacuum region, as shown in Fig. \ref{fig:fig2}(b).  Close to the nanorod, the larger gradient in electrostatic potential induces a phase that varies within the width of the probes, resulting in a loss of fringe visibility. 

We use a $1/r$ potential to approximate the nonlinear monotonically decreasing behavior that is expected surrounding a charged vertical nanorod (Fig. \ref{fig:fig2}(c,d)). In the experimental interferometric BF images there is elongation in the probe separation direction that is not shown in this simple model, but this can be explained by an increased surface charge on the nanorod when the higher order probes ($m\neq\pm1$) are incident on the specimen. Including a Gaussian background to the $1/r$ potential elongated in the probe separation direction to account for this increase in surface charge creates an interference pattern that is qualitatively consistent with the experimental images (Fig. \ref{fig:fig2}(e,f)). 

This initial demonstration shows that the 2GeMZI is sensitive to differences in electrostatic potentials at the locations of the two probes: spatially varying electrostatic potentials impart a phase to the specimen probes and the phase difference modulates the intensity at the BF detector. With moderate improvements, the real-time interferometric BF images can provide nanoscale maps of the static projected potential with higher phase resolution.
Some challenges still need to be overcome to achieve quantitative real-time potential mapping, especially for determining static charge distributions. First, is to ensure the static charge is independent of the scanning probes which can be accomplished with adequately grounded, conductive materials. Second, is to limit the samples of interest such that the spatial extent of the potential does not extend over multiple probes. This ensures there is a reference probe and a measurement probe with a phase difference is directly proportional to the projected potential. This second point can be relaxed with a careful analysis and full understanding of the accumulated relative phases between all the probes as they are scanned into the potential. Third, smaller probe sizes must be used to probe potentials with large spatial gradients.

\section{Quantitative Phase Imaging of a Latex Nanoparticle}

\begin{figure}[h!]
\begin{center}
    \includegraphics[width=3in]{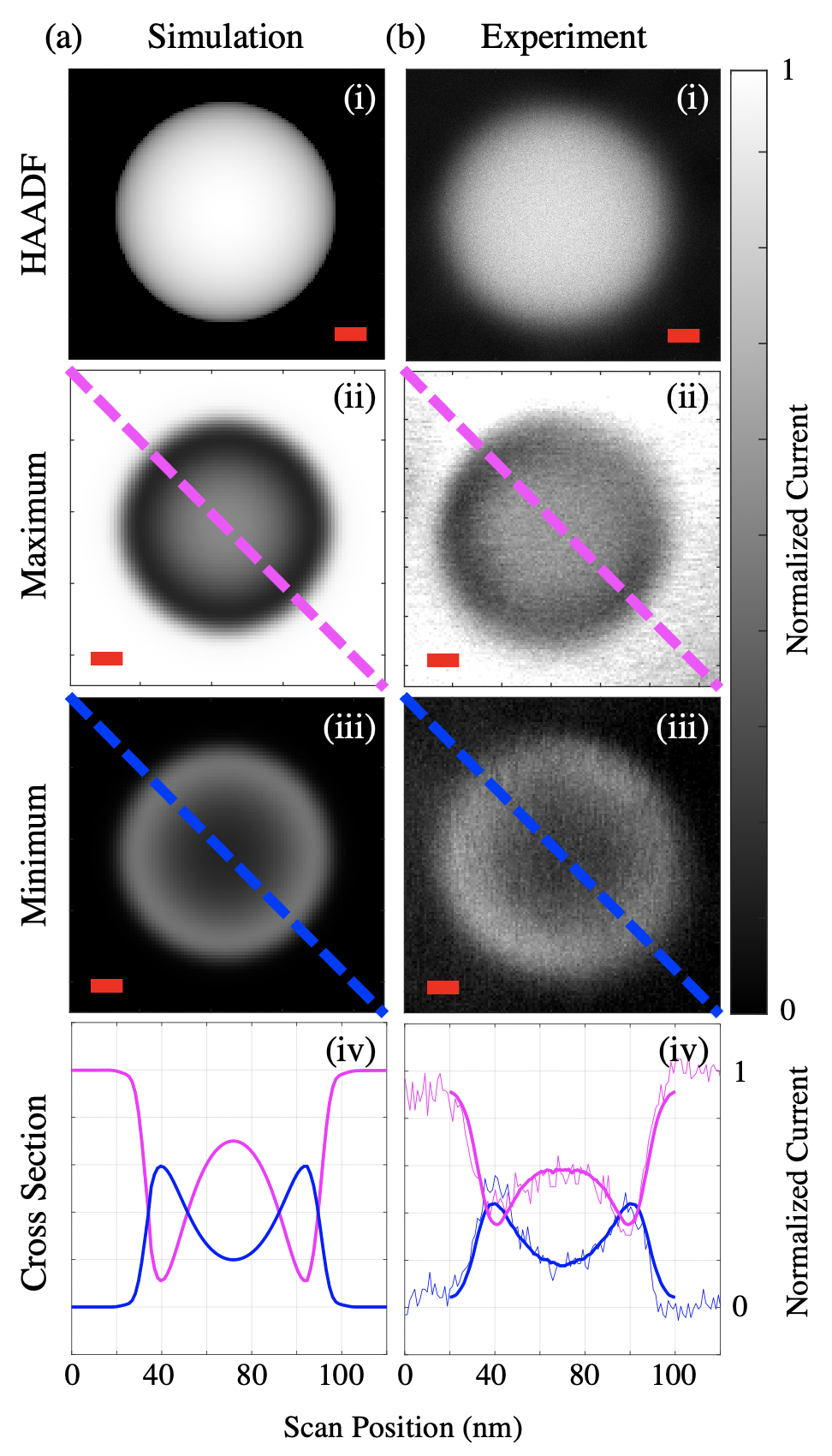}
    \caption{HAADF and 2GeMZI BF image scans of a latex nanoparticle both from (a) simulation and (b) experiment. The rows display (i) the HAADF image, the 2GeMZI image aligned at the maximally (ii) constructive (magenta) and (iii) destructive (blue) interferometer output, and (iv) the respective line profiles. The experimental cross section (b(iv)) shows a raw cross section (light line) and the radially averaged signal (weighted line). All scalebars are 10 nm.}
    \label{fig:Tethers}
\end{center}
\end{figure}

To demonstrate an example of quantitative phase imaging with the 2GeMZI, we imaged polystyrene latex spherical nanoparticles on a suspended single layer of graphene. Polystyrene latex has a well-characterized mean inner potential, $V_{\text{mip}}^{\text{latex}} = 8.5\pm0.7 $ volts \cite{wang_measurement_1998}, and small amplitude decay coefficient, $\gamma^\text{latex}\approx0$. We use a nanosphere with a diameter of 60 nm. The ratio of probe size and separation is fixed, but can be simultaneously tuned by changing the lens magnification settings. Here the 2GeMZI was tuned for a probe separation of $92 \pm 2\, \text{nm}$ and a focused beam width of approximately 5 nm such that one of the two 2GeMZI probes could be scanned through the nanosphere while the other passed through uniform graphene in the specimen plane (Fig. \ref{fig:Tethers}). The phase imparted by the graphene is expected to be about 45 mrad \cite{cooper_atomic_2014}. Individual atoms are not resolvable at the resolution in this initial demonstration, so we treat the sample as a homogeneous latex sphere with a small uniform phase background. Another benefit of the graphene substrate is that it efficiently alleviates charge, allowing us to disregard extraneous static fields due to sample charging and only consider the mean inner potential from the latex as the source of the probe phase shift. Due to the size of the nanoparticle in comparison to the large probe separation, the elimination of electrostatic fields, and the negligible decay coefficient of latex, we can assume the phase difference between the two probes is
\begin{equation}
    \begin{split}
        \varphi(\bk_s,\bx'_0)=& \,\,-2\bk_0\cdot\bx_0' +\sigma V_{\text{mip}}^{\text{latex}}t_{\text{sph}}(\bk_s).\label{eq:phase}
    \end{split}
\end{equation}
The first term, $2\bk_0\cdot\bx_0'$, is due to the interferometer alignment and the last, $\sigma V_{\text{mip}}^{\text{latex}}t_{\text{sph}}(\bk_s)$, is the phase accumulated by the probe passing through the sphere of projected thickness $t_{\text{sph}}(\bk_s)$ at the scan location $\bk_s$. The latex spheres were interferometrically imaged by scanning the probes over a 100x100 $\text{nm}^2$ scan region while the 0$^{\text{th}}$ order interferometer output was recorded by the BF detector.

\begin{figure}[h!]
\begin{center}
    \includegraphics[width=3in]{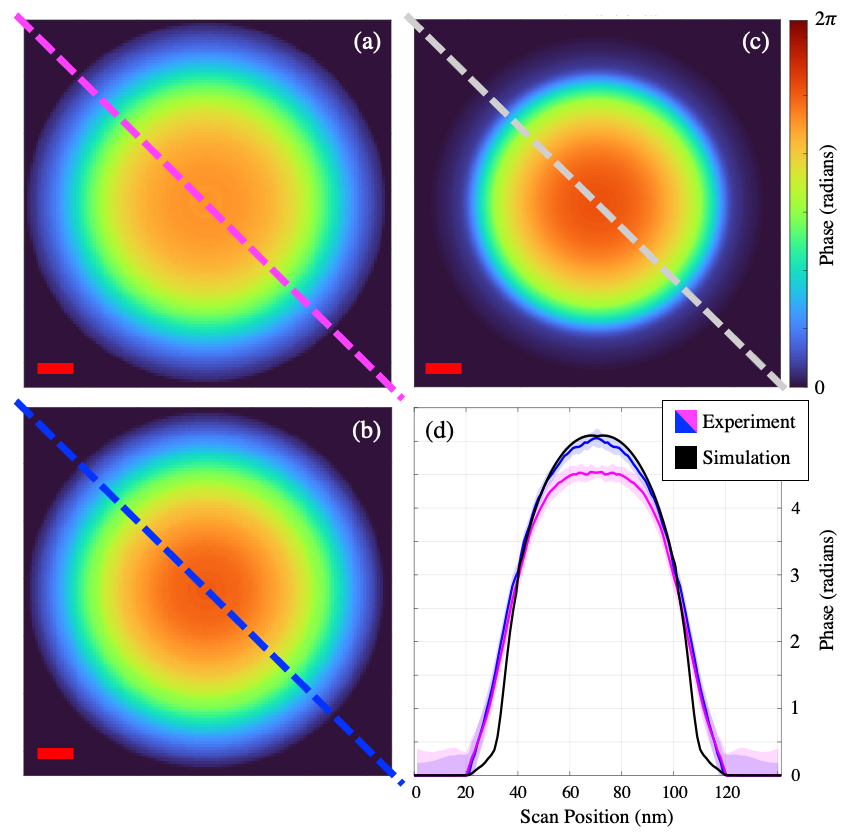}
    \caption{Reconstructed azimuthally averaged phase images of a latex nanoparticle from the raw 2GeMZI BF images with (a) constructive alignment (magenta) and (b) destructive alignment (blue). The simulated outcome (c) is also shown. (d) The experimental and simulated cross-sections of the reconstructed particle phase with shaded regions to show the error. All scale bars are $10\,\text{nm}$.}
    \label{fig:Tethers2}
\end{center}
\end{figure}

We recovered the phase image of the $60 \text{nm}$ diameter latex nanoparticle from two interferometric BF images, one with the interferometer initially aligned for maximally constructive 0$^{\text{th}}$ order output, $2\bk_0\cdot\bx_0'=\pi$ (Fig.~\ref{fig:Tethers}(ii)) and one aligned for destructive output $2\bk_0\cdot\bx_0'=0$ (Fig.~\ref{fig:Tethers}(iii)). To map each pixel's intensity onto a phase, we first find the center of the nanoparticle and take a radial average of the intensity to exploit the particle's symmetry. This provides an intensity line profile across the nanoparticle and graphene from the constructive and destructive interferometric images (Fig. \ref{fig:Tethers}(iv)). Exploiting the phase continuity of the spherical nanoparticle, we note that the phase should be monotonically increasing from the graphene substrate to the center of the nanoparticle. We set the phase of the graphene substrate to zero, as a reference. Using the co-sinusoidal relation between phase and intensity found in Equation (\ref{eq:intout}), we map the intensity profile to a phase profile. From these radial phase profiles, we reconstruct a phase image of the particle, as shown in Fig. \ref{fig:Tethers2}.

Using the experimental Fourier coefficients which define each grating, a 5 nm spot size, 100 nm probe separation and assuming the nanoparticle is a perfectly spherical phase object, we simulate the expected intensity output in the BF detector when the positive first order probe interacts with the sample for both alignment schemes (Fig. \ref{fig:Tethers}(a)). The simulated intensity profile is then mapped to phase using the same mechanism as described above. The experimental phase profiles undershoot the simulated results (Fig.~\ref{fig:Tethers2}(d)) because the nanoparticle's amplitude decay coefficient is non-negligible. The discrepancy between the constructively and destructively interfering experimental results can also be attributed to the amplitude decay. Decoupling the amplitude loss and imparted phase is a subject of ongoing work. The experimental phase profile of the nanosphere is slightly wider than the simulated results, possibly due to induced charging effects.

The 2GeMZI achieves qualitative real-time phase imaging and quantitative phase recovery with a phase resolution of $\sigma_{ph} = 240\, \text{mrad}$. The phase precision could be improved with enhanced gratings, a smaller probe size, and longer exposure times with efficient charge alleviation. The spatial resolution of the 2GeMZI is limited by the focused probe width, which is tunable by selecting different convergence angles using the lens system. Since the holographic grating is used as both a beamsplitter and a beam-defining aperture, the ratio of the maximum separation between probes to the width of those probes is constant and equal to the number of lines in the grating. Considering grating-based phase imaging has previously achieved 30 mrad phase and sub-nanometer spatial resolutions \cite{yasin_probing_2018}, there is promise for phase and spatial resolution improvements of this grating-based technique.

\section{Conclusion}

We constructed a scanning, path-separated, two-grating Mach-Zehnder interferometer (2GeMZI) for free electrons by employing two nanofabricated diffraction grating holograms in a conventional TEM. Although each figure of merit can be tuned or improved, the initial implementation shown here has an interference contrast of $82\%$, a path separation of up to 1 $\mu$m, a demonstrated phase resolution of approximately $\sigma_{ph} = 240\, \text{mrad}$, and an output current on the order of tens of picoamps. We qualitatively showed that the 2GeMZI is sensitive to phase shifts due to electrostatic potentials in vacuum by imaging the potential differences around both grounded and insulated silver vertical nanorods. We then quantitatively recovered the phase of a polysterene latex nanoparticle on graphene. The 2GeMZI is particularly impactful in free electron interferometry due to its tunable probe separation, the accessibility of individual paths, the ability to arbitrarily apply phase shifts between separate paths, its scanning capabilities, and the real-time phase information at the nanoscale. 

With incremental improvements in grating beamsplitters and detectors, the 2GeMZI could be used for interaction-free electron imaging \cite{putnam_noninvasive_2009,kruit_designs_2016}, low-dose STEM imaging \cite{ophus_efficient_2016,agarwal_reduced_2019}, nanoscale magnetic imaging \cite{greenberg_magnetic_2020}, fundamental quantum physics experiments such as the Aharnov-Bohm effect \cite{tonomura_evidence_1986,caprez_macroscopic_2007}, and furthering decoherence theory \cite{schattschneider_entanglement_2018,beierle_experimental_2018}. Subjects of ongoing work are decoupling the imparted phase and amplitude loss, enhancing the contrast at the detector, and improving the spatial and phase resolution. Due to the flexible design and broad applications, the 2GeMZI is uniquely positioned in electron microscopy to open doors to sub-nanometer electron interferometry and low-dose, high-resolution microscopy.

\section*{Acknowledgements}
We thank Joshua Razink and Valerie Brogden for TEM and SEM/FIB instrument support. This work was supported by the NSF Grant No. 1607733. CWJ was supported by the NSF GRFP Grant No. 1309047. 

\section*{Appendix A: Vertical Nanorod Fabrication}

Using a Ga$^+$ FIB operated at 30 keV with a 7.7 pA beam current, we fabricated vertical silver nanorods on a nitride cantilever from a 100 nm thick silver film thermally deposited on a 50 nm thick Si$_3$N$_4$ membrane in the following steps: 
\begin{enumerate}[(i)]
    \item Mill completely through silver and nitride forming cylindrical silver bead along nitride tether.
    \item Mill only through silver defining bottom edge of nanorod, optionally leaving a small lead of silver between the rod and the film.
    \item Mill completely through silver and nitride defining top edge of rod and nitride cantilever.
    \item Flip membrane over and raster FIB over the bare nitride section of cantilever to induce bending until the nanorod is vertical, normal to the silver film and nitride membrane \cite{cui_ion-beam-induced_2013}.
\end{enumerate}

These nanorods were fabricated with a clear vacuum region around the rods for easy access for imaging in the 2GeMZI. One of the two nanorods was given a small lead to ground to the rest of the silver film, while the other was electrically insulated by removing all of the silver between the nanorod and the film. An illustration of the fabrication process can be seen in Figure \ref{fig:fig4}(a) accompanied by scanning electron microscopy (SEM) micrographs displaying the sample geometry and orientation, Fig. \ref{fig:fig4}(b-d).

\begin{center}
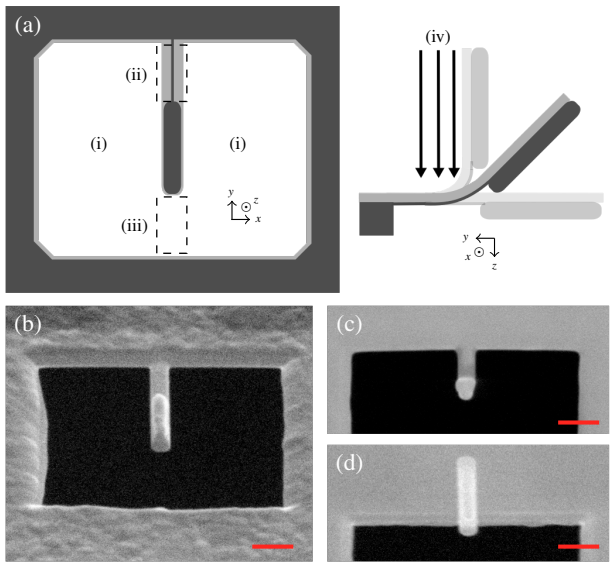
\begin{figure}[h]
\begin{tikzpicture}
    \node[inner sep=0pt] at (0,0)
        {\includegraphics[width=3.3in]{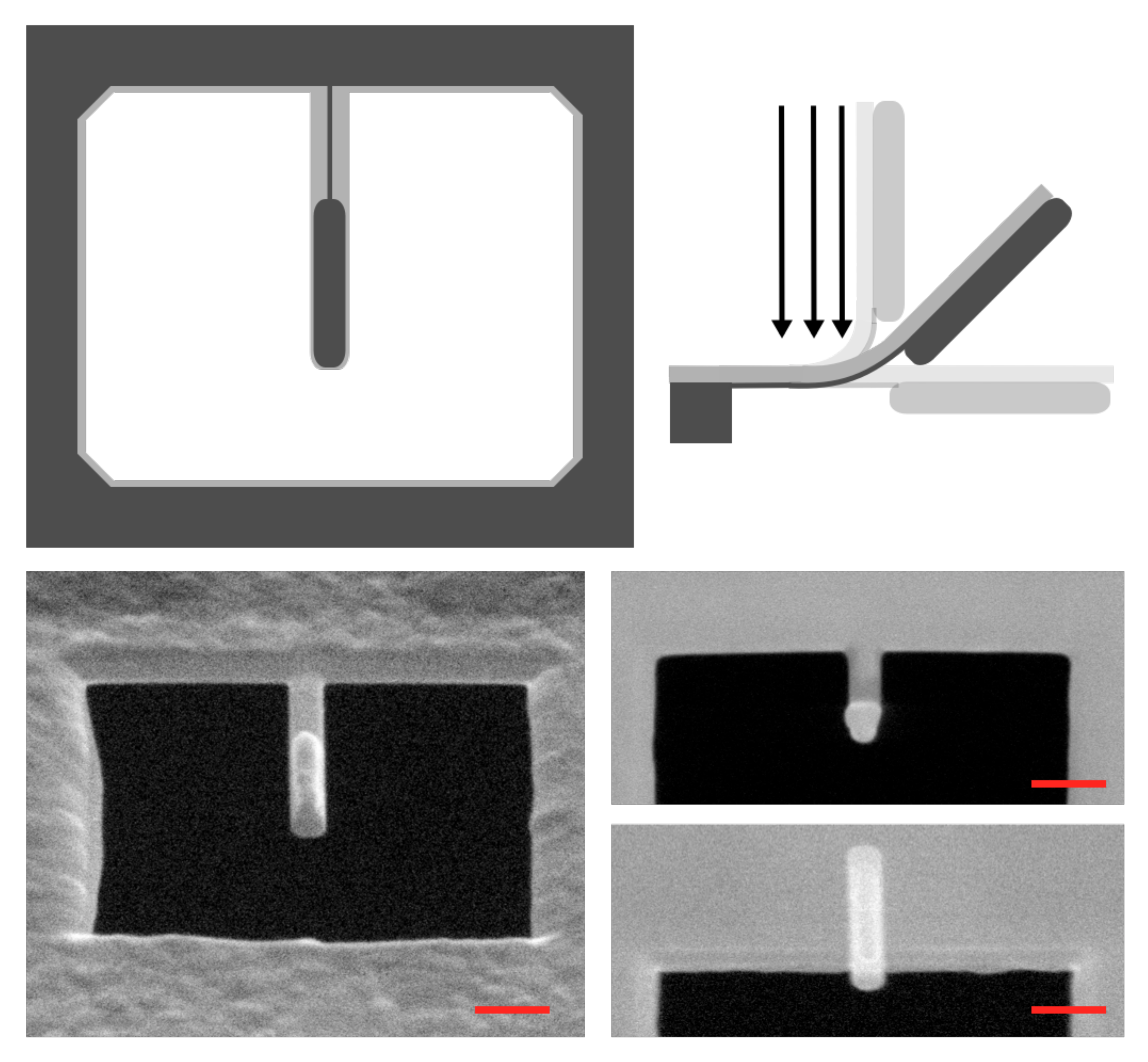}};
    \draw[->,line width=0.15mm] (-1,6.8-5.95) -- (-0.75,6.8-5.95);
    \draw[->,line width=0.15mm] (-0.99,6.8-5.95) -- (-0.99,7.05-5.95);
    \node[scale=0.5,inner sep=0pt] at (-0.8,6.97-5.95)
        {$\boldsymbol{\odot}$};
    \node[scale=0.5,inner sep=0pt] at (-0.64,6.8-5.95)
        {$x$};
    \node[scale=0.5,inner sep=0pt] at (-1,7.14-5.95)
        {$y$};
    \node[scale=0.5,inner sep=0pt] at (-0.67,7.05-5.95)
        {$z$};
    \draw[->,line width=0.15mm] (2.5,6.55-5.95) -- (2.25,6.55-5.95);
    \draw[->,line width=0.15mm] (2.5,6.55-5.95) -- (2.5,6.3-5.95);
    \node[scale=0.5,inner sep=0pt] at (2.3,6.37-5.95)
        {$\boldsymbol{\odot}$};
    \node[scale=0.5,inner sep=0pt] at (2.15,6.3-5.95)
        {$x$};
    \node[scale=0.5,inner sep=0pt] at (2.12,6.55-5.95)
        {$y$};
    \node[scale=0.5,inner sep=0pt] at (2.5,6.175-5.95)
        {$z$};
    \draw[dashed,line width=0.5] (-1.9-0.09,6.35-5.95) rectangle (-1.5-0.09,7.1-5.95);
    \draw[dashed,line width=0.5] (-1.9-0.09,8.3-5.88) rectangle (-1.5-0.09,9.05-5.88);
    \node[scale=0.75,inner sep=0pt] at (-2.6-0.16,7.8-5.95)
        {(i)};
    \node[scale=0.75,inner sep=0pt] at (-0.75-0.16,7.8-5.95)
        {(i)};
    \node[scale=0.75,inner sep=0pt] at (-2.1-0.16,8.65-5.88)
        {(ii)};
    \node[scale=0.75,inner sep=0pt] at (-2.125-0.16,6.7-5.95)
        {(iii)};
    \node[scale=0.75,inner sep=0pt] at (1.75,9.15-5.88)
        {(iv)};
    \node[inner sep=0pt] at (-3.7,3.425)
        {\textcolor{white}{(a)}};
    \node[inner sep=0pt] at (-3.7,-0.55)
        {\textcolor{white}{(b)}};
    \node[inner sep=0pt] at (0.575,-0.55)
        {\textcolor{white}{(c)}};
    \node[inner sep=0pt] at (0.575,-2.4)
        {\textcolor{white}{(d)}};
\end{tikzpicture}
\caption{\label{fig:fig4} (a) Vertical silver nanorod FIB nanofabrication steps (i-iv). (b) SEM micrograph at 52$^\circ$ tilt after fabrication steps (i-iii). SEM micrographs after fabrication steps (i-iv) imaging from bottom of membrane at (c) 0$^\circ$ tilt and (d) 52$^\circ$ tilt. All scale bars are 200 nm.}
\end{figure}
\end{center}

\bibliography{mybib}

\end{document}